\documentclass[12pt,preprint]{aastex}

\shorttitle{The Stellar Content of NGC~2685 and NGC~4650A}
\shortauthors{Karataeva et al.}

\begin{document}

\title{The Stellar Content of the Polar Rings in the Galaxies
NGC~2685 and NGC~4650A\thanks{
Based on observations made with the NASA/ESO Hubble
Space Telescope, obtained from the
Space Telescope Science Institute, which is operated by the Association of
Universities for Research in Astronomy, Inc., under NASA contract NAS5-26555.
}}

\author{G.M.\,Karataeva\altaffilmark{1}, I.O.\,Drozdovsky\altaffilmark{2},
V.A.\,Hagen-Thorn\altaffilmark{1} and  V.A.\,Yakovleva\altaffilmark{1}}
\affil{Astronomical Institute, St.Petersburg State University,
 Universitetskii pr., 28, Petrodvoretz, St.Petersburg, 198504, Russia}

\author{N.A.\,Tikhonov\altaffilmark{3} and O.A.\,Galazutdinova\altaffilmark{3}}
\affil{Special Astrophysical Observatory, N.Arkhyz, Karachai-Circassian Rep.,
369167, Russia}

\altaffiltext{1}{Isaac Newton Institute of Chile, St.Petersburg Branch}
\altaffiltext{2}{Visiting Astronomer, SIRTF Science Center, California
 Institute of Technology, MS 220-6 Pasadena, CA 91125, USA}
\altaffiltext{3}{Isaac Newton Institute of Chile, SAO Branch}

\begin{abstract}
We present the results of stellar photometry of polar-ring galaxies
NGC~2685 and NGC~4650A,
using the archival data obtained with the Hubble Space Telescope's
Wide Field Planetary Camera 2. Polar rings of these
galaxies were resolved into $\sim 800$ and
$\sim 430$ stellar objects in the $B$, $V$ and $I_{\rm c}$ bands,
considerable part of which are blue supergiants located in the young stellar
complexes. The stellar features in the CM-diagrams are best represented by
isochrones with metallicity Z = 0.008. The process of star formation in the
polar rings of both galaxies was continuous and the age of the youngest
detected stars is about 9 Myr for NGC 2685 and 6.5 Myr for NGC 4650A.
\end{abstract}

\keywords{galaxies: individual: (NGC 2685, NGC4650A) --- galaxies:
peculiar --- galaxies: starburst --- galaxies: stellar content}

\section{Introduction}

Polar-ring galaxies (PRGs) represent a rare class of peculiar objects
containing a ring or annulus of gas, stars and dust orbiting in
a plane nearly perpendicular to the equatorial plane of the host galaxy.
This unique geometry of PRGs gives an ideal opportunity to investigate the
shape of their 3D gravitational potential.
Typically, the host galaxy looks like a galaxy of early morphological type
but with some peculiarities in color and brightness distribution;
sometimes these properties let this component more similar to a late-type
object rather than to an S0 galaxy \citep[e.g.][]{arnab95,iodice02}. Recently
polar rings (PRs) have also been found around spiral galaxies NGC~660
\citep{vanDr95} and UGC~5600 \citep{narka01}. The main properties of polar
rings (compiled by \citet{bourncomb03}) are similar to those of  spiral
arms:  blue color, plenty of gas and dust, prominent star forming regions,
and disk-like rotation.

According to the most popular
point of view, PRGs are the result of galaxy interaction, ranging from
simple gas accretion \citep[e.g.][]{schww83} to complete merger
\citep{bekki97}. Both hypotheses are investigated in detail by
\citet{bourncomb03}. The authors conclude that both scenarios may explain
the PRGs formation but the accretion one is preferable because some
predictions relative to the merging scenario (such as the stellar halo around
the polar ring) are not observed.
Alternatively, polar rings can represent the delayed inflows of primordial
gas \citep[e.g.][]{toom77,curir94}.

Intensive studies of this strange objects have been started after
publication of the atlas and catalogue of PRGs and candidates for
these objects \citep{whit90}, though strong emission lines in the spectra
of PRs were detected before \citep[e.g.][]{schgun78,schww83}.
The presence of H{\sc ii}-regions in the PRs of some nearby galaxies,
such as NGC~2685 \citep{makresh89,eskrid97} and NGC~4650A \citep{eskrid99},
the optical colors and H$\alpha$ luminosities of the rings
\citep{reshcomb94} as well as high IR-fluxes of PRGs \citep{richt94}
suggest a presence of active star formation in PRs.

We present here a deep optical study based on {\sl HST\/} WFPC2 archival data
of the resolved stellar populations of two nearby PRGs,
NGC~2685 (D$\sim12.5$~Mpc), and NGC~4650A (D$\sim35$~Mpc)
 (H$_{\rm 0}$~=75~km/s/Mpc).
Multi-color stellar photometry provides the opportunity to
study directly the resolved stellar populations of different masses, ages
and chemical abundances. The analysis of the stellar distribution on
Color-Magnitude Diagrams (CMDs) is the most powerful way to investigate
population
fractions and their spatial variations, to estimate the galaxy distance,
and to provide clues to its star formation history \citep[e.g.][]{tolstoy99}.

Both galaxies were included
by \citet{whit90} in the group of most probable PRGs, 'A', having
two well-established orthogonal kinematic systems on the basis of optical
\citep{sersag72,schgun78,schech84} and radio-H{\sc i} observations
\citep{shane80}. The  important difference between these galaxies is
that the polar ring of NGC~2685 is comparable in size to the host galaxy,
while NGC~4650A has a polar ring extended out to more than two radii of the
host galaxy.

\begin{figure}
\plotone{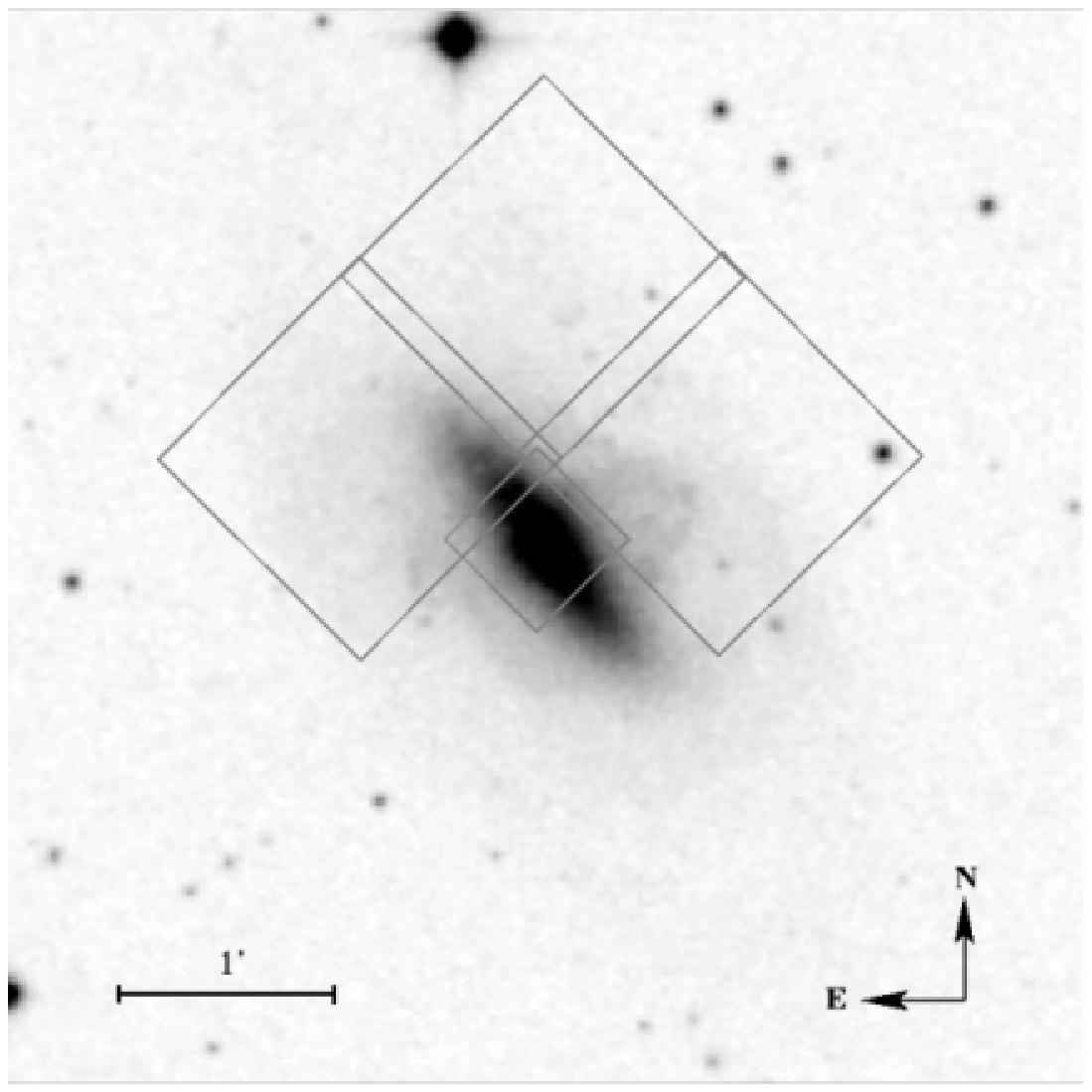}
\caption{DSS $5\arcmin\times5\arcmin$ image of NGC~2685 with
WFPC2 footprint overlaid.
\label{f:fig1}}
\end{figure}

NGC~2685 (Arp~336, also known as "The Helix galaxy'' and "The Spindle'')
has long been known as an unusual object \citep{sandage61}.
Fig.~\ref{f:fig1} presents its image taken from the
The Digitized Sky Survey.
\citet{sandage61} pointed to the singularity of NGC~2685, and
later \citet{Arp66} included the galaxy in his atlas of peculiar galaxies.
The main body of the galaxy having a spindle-like shape is twisted by
luminous filaments (rings) traced by dark strips
on the central galaxy. Deep images (see \citet{sandage61}) reveal an outer
extended low luminosity elliptical structure (outer ring) with position angle
close to that of the galaxy major axis. The combination of photometric
\citep{popov83,makresh89} and spectroscopic \citep{schgun78} data shows
that NGC~2685 is most likely an edge-on S0 galaxy with rings (helices)
rotating approximately perpendicular to the main optical disk.
The H{\sc i} map \citep{shane80,schinnerer02} also shows an extended
structure coincided with outer ring. One small, 16th
magnitude galaxy lies within 10\arcmin (36~kpc in projection) of NGC~2685 and
two more faint, small galaxies are within 20\arcmin \citep{richt94}.
If they are not background objects,then the gas capture during the galactic
interaction might produce such a peculiar morphology.

\begin{figure}
\plotone{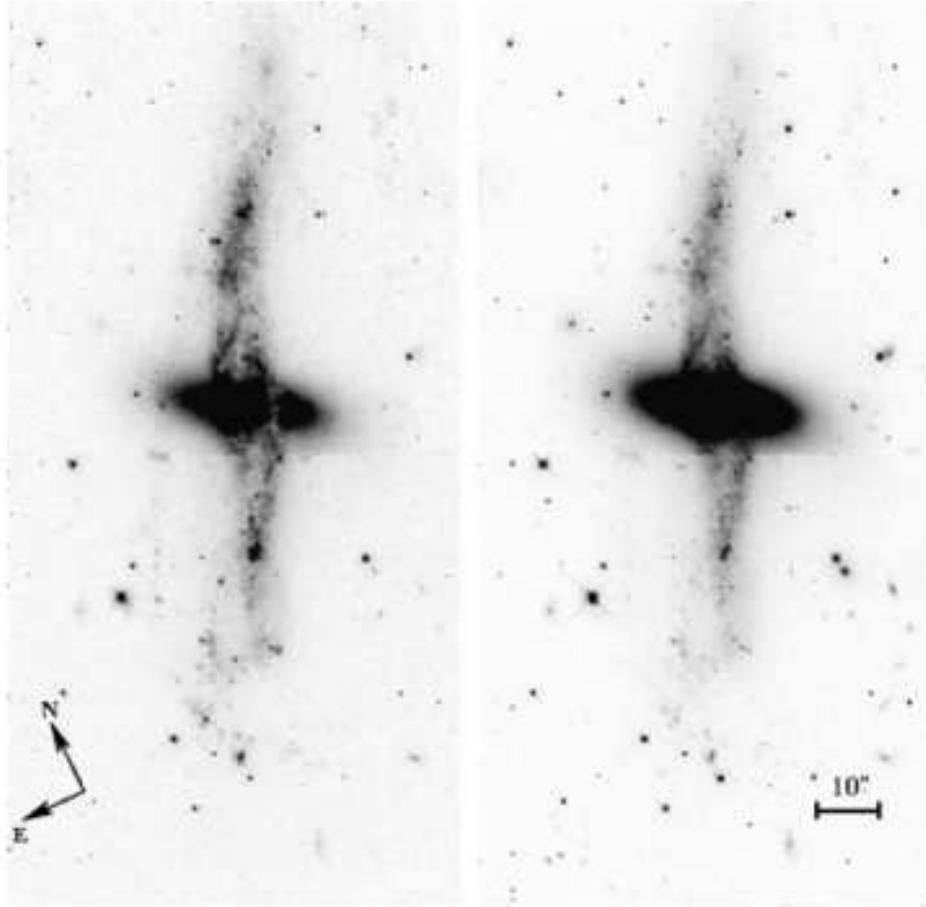}
\caption{WFPC2 images of NGC~4650A (chips WF3 \& 4).
{\em Left},  the $F450W$ band image; {\em right}, the $F814W$ band image.
The blue color of the ring is evident.
\label{f:fig2}}
\end{figure}

Galaxy NGC~4650A (Fig.~\ref{f:fig2}) is a prototype of PRGs with
extraordinary extended polar ring \citep{sers67}. Its structure was
investigated in detail in many works \citep[e.g.][]{gallagher02,iodice02}.
As in the case of NGC~2685 the central galaxy of NGC~4650A looks like
an edge-on S0 galaxy but just this galaxy has peculiar brightness
distribution and colors.
The ring, which is about $2-3$ times the size of the central disk, is
inclined with $\sim100\degr$ angle to the galaxy's major axis.
The dark strip of absorption is visible at the place where ring is
projected on the central galaxy. The ring itself has a complex structure:
in the central part it is inclined to the line of sight and
both parts are visible; howere, at the distance of $30\arcsec$, they
supposedly overlap when the ring become seen as edge-on.
Some of separate filaments are visible in the outer regions of the ring.
According to \citet{arnab97} the ring demonstrates some signs of spiral
structure.

Data available in the literature concerning the color of NGC~2685 polar ring
are inconsistent. In \citep{popov83,makresh89} it has been shown on the basis
of $UBV$ surface photometry that a color of the ring is typical for the
spiral arms of $Sb-Sc$ galaxies. \citet{pel93} found later that rings
color is close to that of central galaxy, i.e. is red. However, this result
is unclear due to uncertainty of transformation of their $J$ and $F$ mag
to the standard Johnson $BV$ magnitude photometric system without taking into
account the large radiation contribution of the emission lines.

The color of NGC~4650A polar ring is blue: $B-V$~$\sim -0\fm1$, according
to \citet{schech84} or $B-V$~$\sim 0\fm2 \div 0\fm3$, according to
\citet{iodice02}.

Both galaxies were observed in the CO line $J=2-1$
\citep{wats94,schinnerer02}. It is established that the emission is
associated solely with the polar rings. \citet{wats94} claimed that in
NGC~4650A the mass of H$_{\rm 2}$ was of
$17\%-35\%$ of the H{\sc i} mass, while in NGC~2685 this value exceeded
H{\sc i} mass. But \citet{schinnerer02} have found that \citet{wats94}
overestimated the H$_{\rm 2}$ mass in NGC~2685 by an order of magnitude.
Nevertheless, both galaxies have sufficient molecular clouds for star forming
activity.

There are some papers containing the claims on stellar populations and
therefore the ages of polar rings \citep{pel93,iodice02,gallagher02}.
For all the cases, results of surface photometry were used.
In  \citet{pel93} the age of the NGC~2685 ring was estimated as
$\sim$ 5-6~Gyr on the basis of its color (however, these data are
uncertain, see above). As for NGC~4650A the data on ring colors show
that the young stellar population certainly exist. The detailed surface
colorimetry of the galaxy was done by \citet{iodice02} and
\citet{gallagher02}. In the first paper the ages of 1-3~Gyr for
the central region of the host galaxy and $< 10^{\rm 8}$ yr  for polar ring
were found, while \citet{gallagher02} claimed that the ages of the host
galaxy and the ring are 3-5~Gyr and $\sim$ 1~Gyr, correspondingly.

A study of different stellar populations of the polar rings, using
their CMDs, will cast light on the origin and evolution of PRGs.
We made here the first attempt to investigate the
stellar population of polar rings in NGC~2685 and NGC~4650A in such a way.

\section{Observations and Basic Data Processing}

We make use of archival HST/WFPC2 data of NGC~2685 and
NGC~4650A. The observations of NGC~2685 were carried out as part of
proposal N~6633  by Marcella Carollo
in January 1999. The data set consists of images in $F450W$,
$F555W$,$F814W$-bands with total exposure times of 2300~sec, 1000~sec
and 730~sec, accordingly. The NGC~4650A data set is based on
the 7500~sec exposures in the $F450W$-band, 4763~sec in
the $F606W$ and 7600~sec in the $F814W$, which were gathered
under proposal N~8399 y Keith Noll in April 1999.

Figure~\ref{f:fig2} presents $F450W$ and $F814W$-band images of
NGC~4650A. The blue color of the ring is evident.
While WF3\&4 chips of WFPC2 cover the most part of NGC~4650A polar ring,
the south part of NGC~2685 and its ring are missing in the WFPC2 data.
Total size of the NGC~2685 according to RC3 is
$4\farcm5 \times 2\farcm3$.
At the distance of NGC2685 $1\arcsec$ corresponds to about 60~pc. For
NGC4650A $1\arcsec$ corresponds to about 170~pc.

The raw frames were processed with the standard WFPC2 pipeline.
The data were extracted from the Archive using the on-the-fly reprocessing
STScI archive system, which reconstructs and calibrates original data
with the latest calibration files, software, and data parameters.
When it was possible we used
the recent version of the {\tt DitherII} package, which performs a
``drizzling'' (a variable-pixel linear for correcting reconstruction)
and corrects for geometric distortion.
The processed frames were then separated into images
for each individual CCD, and trimmed of the vignetted
regions using the boundaries recommended in the WFPC2 Handbook.
The images were combined after cleaning them for bad pixels and
cosmic\--ray events.

The single-star photometry of the images was processed with
{\tt DAOPHOT}/{\tt ALLSTAR}.
To avoid contamination by galactic background emission
single-star photometry was performed on images with
subtracted median-smoothed (with a window of $10\times(FWHM)$)
shape.
The PSF was modeled and evaluated for each chip. The search of the
stellar objects was made using the master frame produced from
all images. The resulting list of stellar coordinates was given to
{\tt ALLSTAR} to perform the photometry in the individual
frames. The necessary corrections were applied to
resulting photometric data including a correction for Charge-Transfer
Inefficiency.
Transformation of \cite{holtzman95} was used to convert instrumental
magnitudes to the standard $B(F450W)$, $V(F555W,F606W)$, and
$I_{\rm c}(F814W)$ system.

$F606W$-band is a wide 'H$\alpha$ continuum' and includes H$\alpha$.
It may influence on the accuracy of the NGC~4650A $V$-band photometry.
Different band star lists were merged requiring a positional source
coincidence better than $0.5 \times FWHM$, or a box size of 2 pixels.

\begin{figure}
\plotone{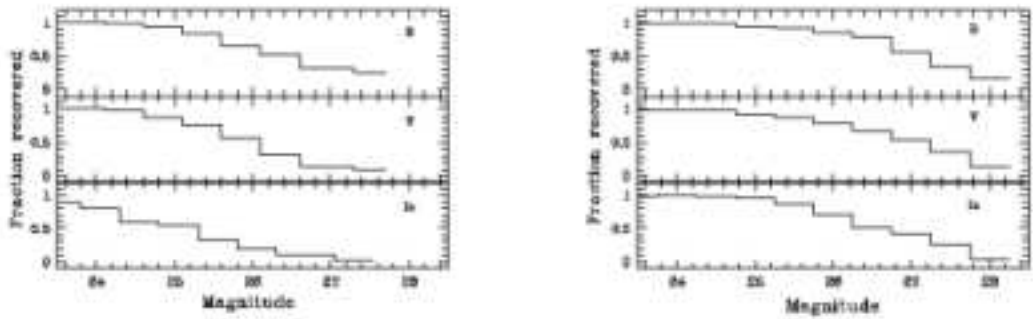}
\caption{NGC~2685 ({\em left}) and NGC~4650A ({\em right}) completeness
levels of the WFPC2 photometry based on artificial star tests.
\label{f:fig3}}
\end{figure}

Completeness test was performed using the usual procedure of artificial
star trials \citep{stetson}. A total of 1,500 artificial stars were added
to the $F450W$ ,$F606W$ ($F555W$) and $F814W$-band frames in several
steps of 150 stars each. These had magnitudes and colors in the range
$18\fm0 \leq $$I_{\rm c}$$ \leq 28\fm0$ and
$0\fm5 \leq $($V$ - $I_{\rm c}$)$ \leq 1\fm5$.
Stars were considered as recovered if they were found in all the bands
with magnitudes not exceeding $0\fm75$ brighter than the initial,
injected ones.
The results of the test are shown in Fig.~\ref{f:fig3}. As one can see
in NGC~2685 the recovery is
practically complete up to m~= $24\fm8$  in $B$-band , m~= $24\fm6$ in
$V$-band and m~= $23\fm5$  in $I_{\rm c}$-band; for NGC~4650A up to m~=
$25\fm6$  in $B$-band, m~= $25\fm2$ in $V$-band and  m~= $25\fm2$ in
$I_{\rm c}$-band. For more faint objects
(within $1\fm0-1\fm5$) the recovering fraction is half as much.

\section{Results}

\begin{figure}
\plotone{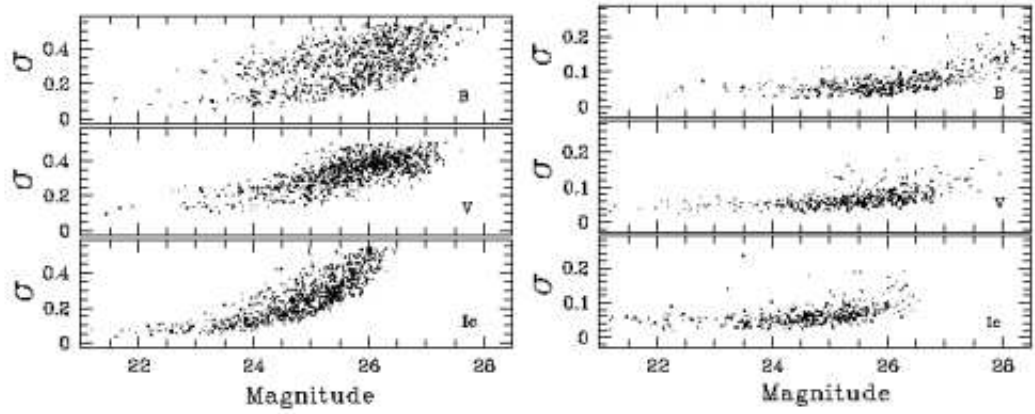}
\caption{NGC~2685 ({\em left}) and NGC~4650A ({\em right}) errors
 of the WFPC2 photometry.
\label{f:fig4}}
\end{figure}

There are $\sim 450$ stellar objects in NGC~4650A and $\sim~800$
in NGC~2685 detected in $B$, $V$ and $I_{\rm c}$ filters with
quality parameters $SHARP$ and $CHI$ (as defined in {\tt ALLSTAR}
in {\tt MIDAS})
in the intervals $-1 \leq SHARP \leq 1$  and $CHI \leq 2$.
Our final lists of stellar objects include objects having angular sizes
$FWHM<0\farcs15$ for NGC~4650A and $FWHM<0\farcs40$ for NGC~2685.
Figure~\ref{f:fig4} illustrates the precision of our photometry.

\begin{figure}
\plotone{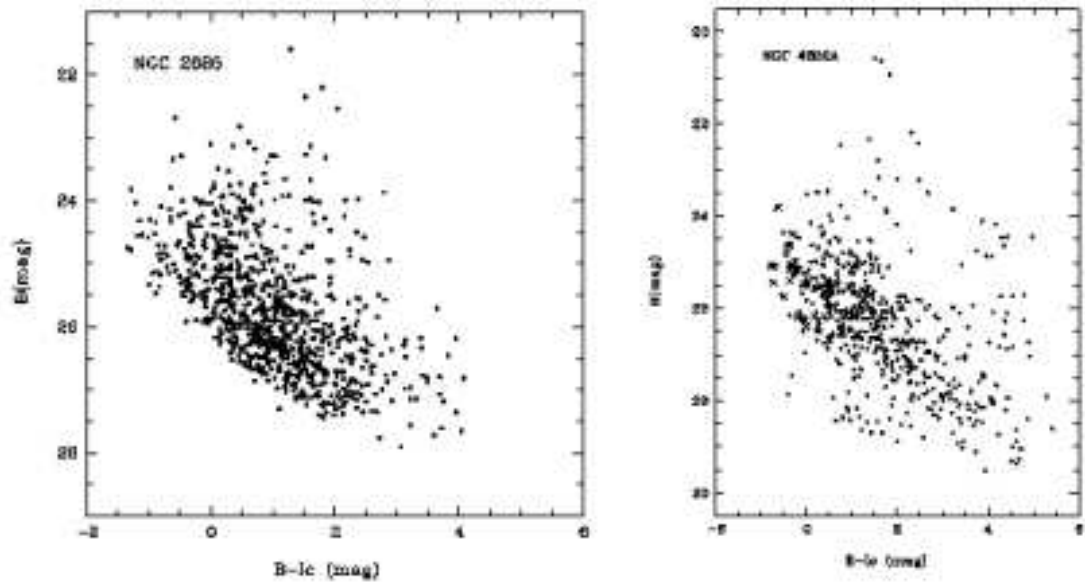}
\caption{color--magnitude diagrams for NGC~2685 ({\em left}) and
NGC~4650A ({\em right}) corrected for absorption in our Galaxy (crosses
are unresolved H{\sc ii}-regions).
\label{f:fig5}}
\end{figure}

The magnitudes and colors of these objects were corrected for extinction in
our galaxy using IRAS/DIRBE map of \cite{schlegel}. With $R_{\rm V}~=~3.1$
 and the extinction law of \cite{cardelli}, galactic foreground extinction for
NGC~4650A is:
 $A_{\rm B} = 0\fm48$, $A_{\rm V} = 0\fm37$,
 and $A_{\rm I_{\rm c}} = 0\fm22$; and for NGC~2685:
 $A_{\rm B} = 0\fm27$, $A_{\rm V} = 0\fm21$, and $A_{\rm I_{\rm c}} = 0\fm12$.
The extinction-corrected data are used for construction of color-
magnitude and color-color diagrams. CMDs ($B$ vs $(B-I_{\rm c})$) for
NGC~2685 and NGC~4650A are given in Figure~\ref{f:fig5}.

Among detected point objects the bulk of which are evidently blue
giants and supergiants the outside objects with dimensions less than
25~pc may be present.
The possible candidates are unresolved double and multiple stars, super star
clusters (SSC) with typical dimensions of 8$\div$10~pc, compact
H{\sc ii}-regions, and more extended
H{\sc ii}-regions with typical dimensions of about 20~pc.

It is difficult to separate multiple stars from single supergiants in our
data, but a portion of systems in which single blue supergiant is not
dominant probably is not high (see below). There are no SSCs among our 
objects. Having mean absolute magnitudes of
M$_{\rm V} = -13^{\rm m}$ \citep{whit93} SSCs
at the distance of NGC~2685 (m-M~= $30\fm5$) will have an apparent
magnitude of $17\fm5$ and $19\fm7$ at the
distance of NGC~4650A (m-M~= $32\fm7$). There are no such bright objects
among those listed. H{\sc ii}-regions can be separated using two-color
diagrams. Since the H${\alpha}$ emission penetrates to the $F606W$-band
the $V-I_{\rm c}$ color for H{\sc ii}-regions proves to be less than one
for blue stars ($V-I_{\rm c} \sim -0\fm35$). Therefore, H{\sc ii}-regions
have bluer color in $V-I_{\rm c}$ than in $B-I_{\rm c}$. As a result all
questionable objects were excluded from the final lists, which contain 430
stars for NGC~4650A and 800 stars for NGC~2685.

\subsection{NGC2685}

\begin{figure}
\plotone{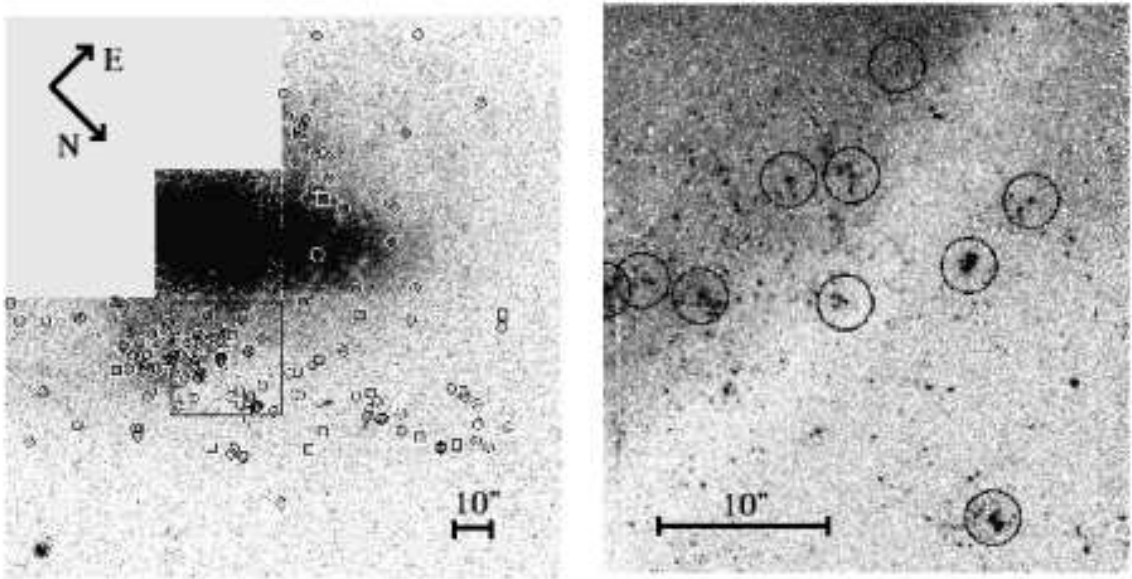}
\caption{WFPC2 image of NGC~2685.
Circles and squares indicate the position of blue and red supergiants,
correspondingly ({\em left}). The candidates to H{\sc ii} regions in
the enlarged part of NGC~2685 polar ring ({\em right}).
\label{f:fig6}}
\end{figure}

In Fig.~\ref{f:fig6},({\em left}) the location of blue (circles) and
red (squares) supergiants of NGC~2685 is shown. It is evident that most
part of both are concentrated to the region of polar rings (though some
objects in the lower-right corner, perhaps, belong to the outer ring).

In the Fig.~\ref{f:fig6},({\em right}) we give in enlarged scale the
map of the
portion of galaxy near polar ring in which circles are the most blue
regions found after dividing frame $F814W$ by $F550W$. All of these
coincide with H{\sc ii}-regions. From comparison Fig.~\ref{f:fig6},
({\em left}) and Fig.~\ref{f:fig6}, ({\em right}) the correlation between
locations of H{\sc ii}-regions and blue supergiants is evident.

As one can see in the Fig.~\ref{f:fig6}, the amount of red supergiants is
quite small. It is noticeable that both
red and blue supergiants tend to gather into stellar associations.
This fact also confirms that the detected stars are red supergiants
of the polar ring rather than foreground Galactic red stars.

In the color-magnitude diagram $B$ vs $(B-I_{\rm c})$ (Fig.~\ref{f:fig5})
the blue limit of the stars is shifted to the red in comparison with
normal position for blue supergiants which can be found from
\citep{bertelli94}.
This fact point to significant intrinsic absorption in polar rings.

Comparison of the blue limit of the stars for the galaxy with that
found for galaxies without intrinsic absorption gives the color-excess for
the galaxy. But blue limit at CMDs is not sharp because of different
absorption for various stars and their different ages and metallicity.
Therefore it is more reliable to use as a compared parameter the color-index
for which the maximum in number of blue stars is observed (this value gives
the mean color index for blue supergiants and may be found after constructing
the corresponding histogram).

For NGC~2685 $(B-I_{\rm c})^{\rm max}~= +0\fm25$; the comparison with galaxies
NGC~959 ($(B-I_{\rm c})^{\rm max}~= -0\fm18$) and HolmII
($(B-I_{\rm c})^{\rm max}~= -0\fm10$)  (our unpublished data; both
galaxies may be considered as galaxies with negligible intrinsic absorption)
 gives  E$_{\rm {(B-I_{\rm c})}}~= 0\fm39$.

There is direct determination of absorption in the
rings obtained from the surface photometry \citep{popov83}. It is known
that the north-east part of the main body of the galaxy is intersected by
four dark strips for which in the absorption was found \citep{popov83}. These
strips are the result of projection on the main body of polar rings which
are luminous outside the galaxy (they simply follows each other). For normal
extinction curve \citep{cardelli} and mean value of absorption in $B$-band
for all four strips one can find E$_{\rm {(B-I_{\rm c})}}~=0\fm35$.

The total absorption in several H{\sc ii}-regions
was obtained by \citet{eskrid97} on the basis of Balmer decrement.
The authors found that absorption $A_{\rm V}$ is essentially different
in different H{\sc ii}-regions (from $0\fm00$ to $0\fm77$). After
correction for galactic extinction it gives
$E_{\rm {(B-I_{\rm c})}} = 0\fm40$ for the most obscured H{\sc ii}-regions.

Two latter estimations do not contradict to the value
E$_{\rm {(B-I_{\rm c})}}~= 0\fm39$ found above. Therefore we accept this.
Since there is no possibility to correct for intrinsic absorption the data
for individual stars we take it into account {\it in average}, reducing for
all stars the color index $B-I_{\rm c}$ by $0\fm39$ and magnitudes $B$, $V$
and $I_{\rm c}$ by $0\fm74, 0\fm56, 0\fm 35$, accordingly.

The resulting M$_{\rm {I_{\rm c}}}$ vs $B-I_{\rm c}$ diagram
(for m-M~= $30\fm5$) is shown in Fig.~\ref{f:fig7}, {\em left}.

\subsection{NGC4650A}

In the CMD for NGC~4650A (Fig.~\ref{f:fig5}) crosses are unresolved
H{\sc ii}-regions selected using two-color diagram. They have been
excluded from the final list.

The location of detected blue stars in the field of NGC~4650A is shown
in  Fig.~\ref{f:fig8}, {\em left}.
The blue objects are evidently concentrated to the polar ring. Note that some
stars follows so called "spiral arms" \citep{arnab97,gallagher02} especially
in their south branch. As one can see in Fig.~\ref{f:fig8}, {\em right}
red stars are distributed randomly without any concentration to the galaxy
and are, therefore, foreground stars belonging to our Galaxy. (In NGC~4650A
we cannot reach the red giants which might consist a halo around polar ring
as the merging scenario demands.)

A comparison of the distribution of blue objects with the H$\alpha$ image of
the galaxy \citep{eskrid99} shows (as in the case of the NGC~2685) a
correlation between location H{\sc ii}-regions and blue stars.

It is difficult to correct for intrinsic absorption in this case.
According to \citet{gallagher02} there is a dust ring in the center
of the galaxy. \citet{iodice02} give the
$A_{\rm B}~= 1\fm54$, $A_{\rm I_{\rm c}}~= 1\fm00$ for the
extinction in the place of the ring projection on the main body. These
values may be considered as an upper limit of intrinsic absorption
in the ring outside the main body in regions of dust ring. But
extreme blue ring colors far off the galactic center show that the
extinction is very small (if any) in these regions. We attempt to
introduce the correction for intrinsic absorption as linear along
the polar ring from the center to periphery but without success
because the scatter in the CMD along $B-I_{\rm c}$ axis has not decrease.
Therefore, we did not apply any correction for intrinsic absorption
keeping in mind that in reality stars at CMD may be located more left.

  The resulting M$_{\rm {I_{\rm c}}}$ vs $B-I_{\rm c}$ diagram
(for m-M~= $32\fm7$) is shown in Fig.~\ref{f:fig7}, {\em right} .

\section{Discussion}

    The important argument that benefits the assumption that most of
the objects are blue supergiants gives the construction of luminosity
functions (LFs) for blue objects in both galaxies.
Luminosity function was calculated for all objects
with  $B-I_{\rm c}< 0\fm8$ for NGC~4650A and $B-I_{\rm c}< 1\fm2$ for
NGC~2685 (taking into account its reddening of
E$_{\rm {(B-I_{\rm c})}}$~= 0\fm39) by counting stars lying inside
a bin of $0\fm5$. The central value was varied in steps of $0\fm25$
to reduce the dependence of our results on the particular choice of bin
center. The results are given in Fig.~\ref{f:fig9}.

  As it was shown by \citet{freedm85} the slope of the luminosity function
for blue supergiants is similar for all spiral galaxies with mean
value of about 0.6. The slope of luminosity function in the bright part
(where the completeness is close to $100\%$) is
$0.62\pm0.01$ for NGC~4650A and $0.61\pm0.02$ for NGC~2685 (dashed lines in
Fig.~\ref{f:fig9}), being  in agreement with \citet{freedm85} value.
Thus the results show that the polar rings contain young blue supergiants
testifying that stars-forming processes are still going there.

\begin{figure}
\plotone{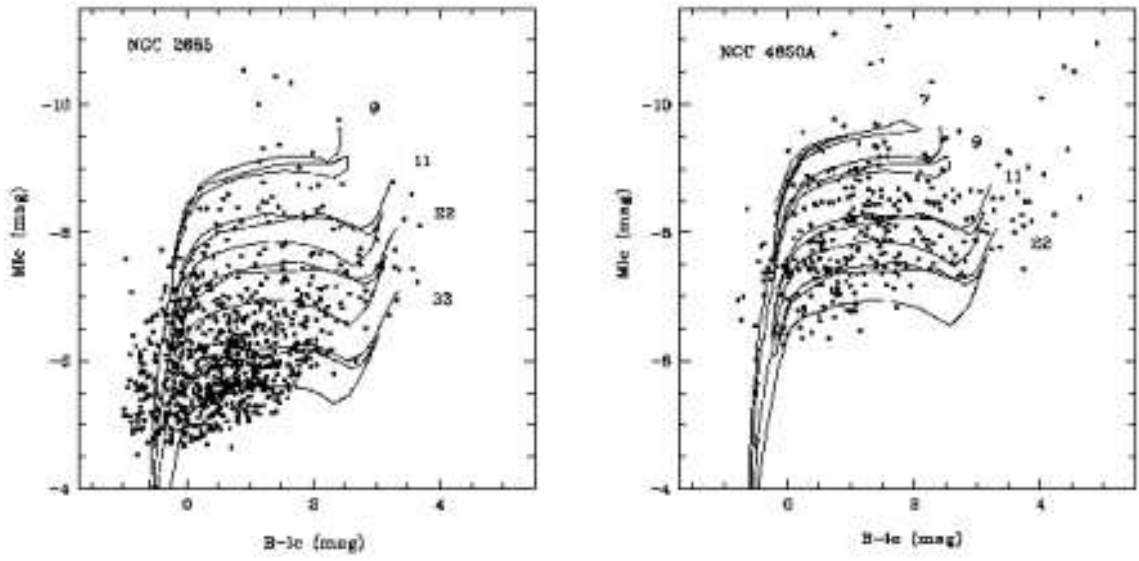}
\caption{M$_{\rm {I_{\rm c}}}$ vs $B-I_{\rm c}$ CMDs of
NGC~2685 ({\it left}) and NGC~4650A ({\it right}).
Stellar isochrones for the metallicity  Z=0.008 from the
Padova library are overplotted for ages from 7~Myr to 33~Myr.
\label{f:fig7}}
\end{figure}

\begin{figure}
\plotone{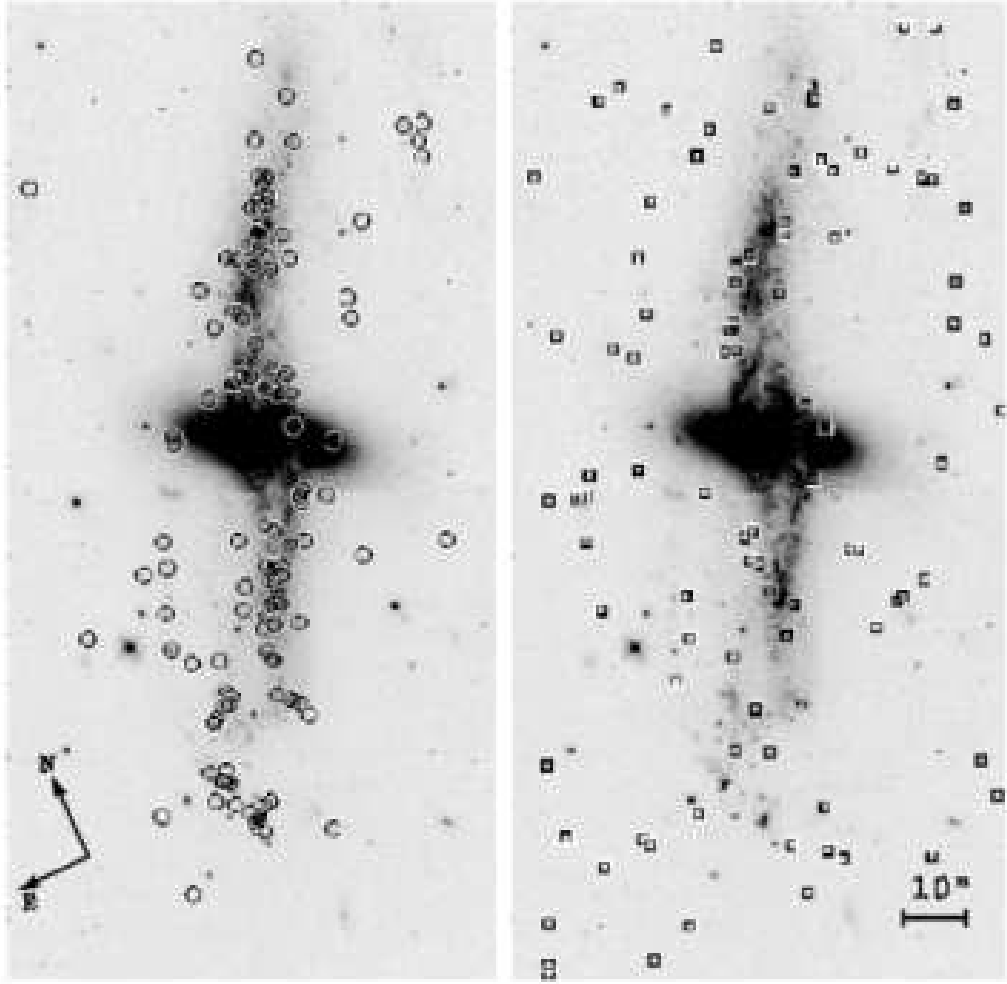}
\caption{Blue supergiants (circles, {\em left}) in NGC~4650A and red stars
 (squares, {\em right}).
\label{f:fig8}}
\end{figure}

\begin{figure}
\plotone{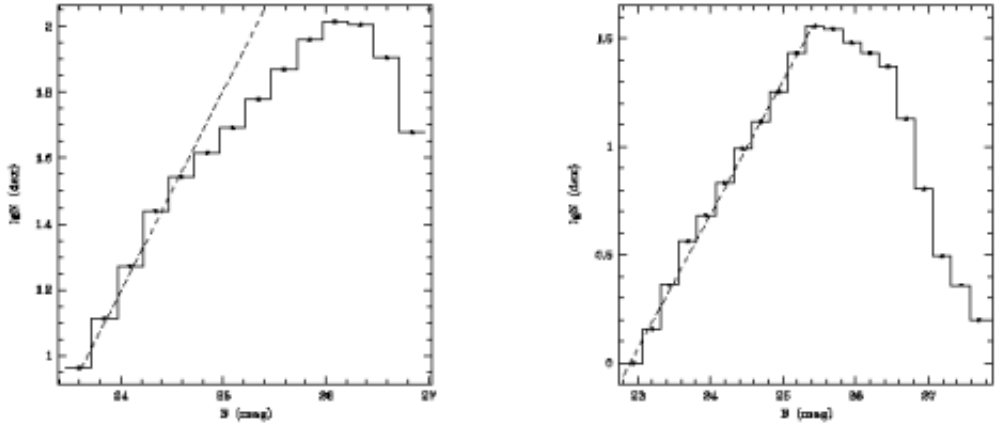}
\caption{Smoothed luminosity function of blue stars (solid line) and
the bright-end slope of the LF (dashed line) of NGC~2685 (left) and
NGC~4650A (right).
\label{f:fig9}}
\end{figure}

   The ``color-absolute magnitude'' diagrams for both
galaxies presented in Fig.~\ref{f:fig7} demonstrate mostly the brightest
and youngest stellar populations. CMD for NGC~2685 is deeper
in comparison with that of NGC 4650A but still does not reach an area of
old/intermediate-age stellar populations, for example,
red giants (M$_{\rm {I_{\rm c}}}\approx -4\fm0$). The discovery of red giants
 (old population) would allow to estimate the age of the rings.

 In Fig.~\ref{f:fig7} the theoretical stellar isochrones from
\citet{bertelli94} are overplotted for the metallicity of
Z~= 0.008 and ages from 7 to 33~Myr. Such a metallicity is usual for irregular
galaxies \citep{garnett02}. Isochrones for this metallicity are in good
agreement with CMDs
observed by us. We do not find any confirmation of high metallicity in
NGC~2685 found by \citet{eskrid97} who used empirical method of its
determination, which strongly depends on accepted intrinsic absorption for
H{\sc ii} regions. The isochrones for higher metallicity (e.g. solar)
certainly contradict with CMD observed.

   From Fig.~\ref{f:fig7} we can infer that process of star formation was
continuous during tens of Myrs, and the last burst of star formation probably
was more recent in NGC~4650A than in NGC~2685.

\section{Conclusions}

    The main result of this work is resolution of the stars in the PRs
of NGC~2685 and NGC~4650A. $B,V,I_{\rm c}$ photometry was made for
several hundreds of stars (mostly blue giants and supergiants) in each
galaxy. Our research is the first one relating to resolvable stellar
population in polar rings.

     The distribution of blue and red supergiants in the field of galaxies
strongly follows the polar rings. The stars tend to gather into stellar
associations which positions are correlated with positions of
H{\sc ii} regions.

     CMDs for both galaxies show that process of star formation is
continuous during tens Myrs, the last burst of star formation being very
recently (6.5~Myr in NGC~4650A and 9~Myr in NGC~2685). Unfortunately, CMDs
for both galaxies are insufficiently deep to claim anything about the old
population
in the rings. But for NGC~2685 the upper limit of red giants branch may be
reached with more deep exposures at HST and planning such observations is
very desirable. These observations might also verify whether a stellar halo
exist around the main galaxy as predicted by \citep{bourncomb03} for merging
scenario of PRGs formation.

\acknowledgements
 This work was supported by RFBR via grants 03-02-16344 and 02-02-16033.

\clearpage

\clearpage 

\clearpage

\clearpage

\clearpage

\clearpage

\clearpage

\clearpage

\clearpage

\end{document}